\documentclass[aps,showpacs,showkeywords,preprintnumbers,nofootinbib, epsf]{revtex4}
\usepackage{graphicx}

\begin{document}

\title{Path and Path Deviation equations of  Fractal Space-Times: A
Brief Introduction}
\author{M. E. Kahil}
\email{kahil@aucegypt.edu} \affiliation{Mathematics Department,
Modern Sciences and Arts University, Giza, Egypt}
\author{T. Harko}
\email{harko@hkucc.hku.hk} \affiliation{Department of Physics and
Center for Theoretical and Computational Physics, The University
of Hong Kong, Pok Fu Lam Road, Hong Kong}

\date{\today }

\begin{abstract}
The idea that the quantum space-time of microphysics may be
fractal everywhere was intensively investigated recently, and
several authors have presented the geodesic equations of different
fractal space - times. In the present work we obtain the geodesic
and the geodesic deviation equations in fractal space-times by
using the Bazanski method. We also extend this approach to obtain
the equations of motion for spinning and spinning charged
particles in the above-mentioned spaces, in a similar way to their
counterparts in Riemannian geometry.
\end{abstract}



\maketitle

\section{Introduction}

The proposal that the space-time in which the evolution of the
microscopic objects takes place may be a fractal has attracted a
lot of attention recently \cite{NoSc84,No94,No98,No05, Ag05,
AgGo06, Go06,No06}. A fractal structure is a manifestation of the
universality of self-organization processes, a result of a
sequence of spontaneous symmetry breaking. In the fractal
space-time model the typical trajectories of quantum particles are
continuous, but non-differentiable, and can be characterized by a
fractal dimension that jumps from $D=1$ at large length scale to
$D=2$ at small length scale. The fractal dimension $D=2$ is the
fractal dimension of the Brownian motion, or, equivalently, of a
Markov-Wiener process. As suggested by Nelson \cite{Ne66}, quantum
mechanics can also be interpreted as assuming that any particle is
subjected to an underlying Brownian motion of unknown origin,
which is described by two (forward and backward) Wiener processes:
when combined together, they wield the complex nature of the wave
function and they transform Newton's equation of dynamics into the
Schrodinger equation.

An alternative way for the description of the motion was
introduced by Bazanski \cite{Ba89}, an approach that has the
advantage of providing both the equations of the geodesics as well
as the geodesic deviation equation. It is the purpose of the
present paper to generalize the Bazanski approach to the case of
the fractal space-times.

\section{Fractal space-times and the Schrodinger equation}

Let us consider a fractal curve $f(x)$ between two points $A$ and
$B$ that is continuous, but nowhere differentiable. Such a curve
has an infinite length. This is a direct consequence of the
Lebesgue theorem, which states that a finite length curve is
almost everywhere (i.e., except a set of points with null
dimension) differentiable \cite{NoSc84, No06}. Another important
property of a fractal curve is that between any two points of the
curve we can get a curve with the same properties as the initial
curve, that is, a continuous, nowhere differentiable and infinite
length curve. Therefore fractal curves are almost self-similar
everywhere.

In the differentiable case the usual definition of the derivative
of a given function are given by
\begin{equation}
\frac{df}{dt}=\lim_{\Delta t\rightarrow +0}\frac{f\left( t+\Delta
t\right) -f(t)}{\Delta t}=\lim_{\Delta t\rightarrow
-0}\frac{f\left( t\right) -f(t-\Delta t)}{\Delta t},
\end{equation}
and one can pass from one definition to the other by the transformation $%
\Delta t\rightarrow -\Delta t$. The differentiable nature of the
space-time implies the local differential (proper) time reflection
invariance. In the non-differentiable case two functions
$df_{+}/dt$ and $df_{-}/dt$ are defined as explicit functions of
$t$ and $dt$,
\begin{equation}
\frac{df_{+}}{dt}=\lim_{\Delta t\rightarrow +0}\frac{f\left(
t+\Delta t,\Delta t\right) -f(t,\Delta t)}{\Delta
t},\frac{df_{-}}{dt}=\lim_{\Delta
t\rightarrow -0}\frac{f\left( t,\Delta t\right) -f(t-\Delta t,\Delta t)}{%
\Delta t},
\end{equation}
with the plus sign corresponding to the forward process, while the
minus sign corresponds to the backward process. In other words,
the non-differentiable nature of the space-times implies the
breaking of the local differential (proper) time reflection
invariance. If we apply the definition of the derivatives to the
coordinate functions, we obtain $dX_{\pm }^{i}=dx_{\pm }^{i}+d\xi
_{\pm }^{i} $, where $dx_{\pm }^{i}$ are the usual classical
variables, and $d\xi _{\pm }^{i}$ are the non-differentiable
variables. By taking the average of these equations we obtain
$\left\langle dX_{\pm }^{i}\right\rangle =\left\langle dx_{\pm
}^{i}\right\rangle $, since $\left\langle d\xi _{\pm
}^{i}\right\rangle =0$. If we denote by
$d\vec{x}_{+}/dt=\vec{v}_{+}$ the
forward speed and by $d\vec{x}_{-}/dt=\vec{v}_{-}$ the backward speed, then $%
\left( \vec{v}_{+}+\vec{v}_{-}\right) /2$ may be considered as the
differentiable (classical speed), while $\left( \vec{v}_{+}-\vec{v}%
_{-}\right) /2$ is the non-differentiable speed. These two
quantities can be
combined in a single quantity if we introduce the complex speed $\vec{V}%
=\delta \vec{x}$, where $\delta
=\left(d_{+}+d_{-}\right)/2dt-i\left(d_{+}-d_{-}\right)/2dt$
\cite{No94,No98,No05, Ag05, AgGo06, Go06, No06}.

By considering that the continuous but non-differentiable curve of
motion is immersed in a three-dimensional space, any function
$f\left( X^{i},t\right) $ can be expanded into a Taylor series as
$df=f\left( X^{i}+dX^{i},t+dt\right) -f\left( X^{i},t\right)
=\left[ (\partial /\partial X^{i})dX^{i}+(\partial /\partial
t)dt\right] f\left( X^{i},t\right) $. By assuming that the mean
values of the function $f$ and of its derivatives coincide with
themselves we obtain $d_{\pm }f/dt=\partial f/\partial
t+\vec{v}_{\pm }\cdot \nabla f_{\pm }$, while the operator $\delta
$ is given by $\delta f/dt=\partial f/\partial t+\vec{V}\cdot
\nabla f$ \cite{No94,No98,No05, Ag05, AgGo06, Go06,Wa99}.

We can now apply the principle of the scale covariance, which
postulates that the passage from classical (differentiable)
mechanics to the non-differentiable mechanics can be realized by
replacing the standard time derivative $df/dt$ by the complex
operator $\delta /dt$. Therefore in a covariant form the equation
of geodesics of the fractal space-time can be written as
\begin{equation}
\frac{\delta \vec{V}}{dt}=\frac{\partial \vec{V}}{\partial
t}+\vec{V}\cdot \nabla \vec{V}=0.
\end{equation}

By considering that the fluid is irrotational, $\nabla \times
\vec{V}=0$, by introducing the complex speed potential $\phi $ so
that $\vec{V}=\nabla \phi $, and by assuming that $\phi =-2iD\ln
\psi $, where $D$ is a constant, the equation of motion of the
fluid takes the form of the Schrodinger equation,
\begin{equation}
D^{2}\Delta \psi +iD\frac{\partial \psi }{\partial t}-U\psi =0,
\end{equation}
where $U=D^{2}\Delta \ln \psi $. $D$ defines the
differential-non-differential transition, that is, the transition
from the explicit scale dependence to scale independence.

In order to study gravitational phenomena in fractal space-times
it is necessary to extend the concept of metric by taking into
account the fluctuating character of the paths. This corresponds
to the passage from Special Scale Relativity to General Scale
Relativity. The line element between two neighboring points in a
fractal geometry can be described as
\begin{equation}
d\tilde{s}^{2}=\tilde{g}_{\mu \nu }dX^{\mu }dX^{\nu
}=\tilde{g}_{\mu \nu }\left( dx^{\mu }+d\xi^{\mu}\right)\left( dx^{\nu }+d\xi^{\nu }\right)  ,
\end{equation}
leading to a generalized metric in a curved fractal space-time of
the form
\begin{equation}
\tilde{g}_{\mu \nu }\left( x,t\right) =g_{\mu \nu }\left(
x,t\right) +\gamma _{\mu \nu }\sqrt{\left( \frac{\lambda
_{c}}{dx^{\mu }}\right) \left( \frac{\lambda _{c}}{
dx^{\nu}}\right) },
\end{equation}
where $\gamma_{\mu \nu}$ is described as the first approximation
in terms of stochastic variables \cite{No06}. Based on this
fluctuating metric one
can obtain the affine connection of the fractal space-time as $\tilde{\Gamma}%
_{jk}^{i}=\Gamma _{jk}^{i}+\chi _{jk}^{i}$, where $\Gamma
_{jk}^{i}$ is the usual Christoffel connection, and $\chi
_{jk}^{i}$ is the fluctuating part.
The mean values of the affine connection satisfy the conditions $%
\left\langle \tilde{\Gamma}_{jk}^{i}\right\rangle =\Gamma _{jk}^{i}$ and $%
\left\langle \chi _{jk}^{i}\right\rangle =0$. Similarly one can
define the
curvature tensor $\tilde{R}_{jkl}^{i}=R_{jkl}^{i}+\Xi _{jkl}^{i}$, so that $%
\left\langle \tilde{R}_{jkl}^{i}\right\rangle =R_{jkl}^{i}$ and $%
\left\langle \Xi _{jkl}^{i}\right\rangle =0$.

\section{The Bazanski approach in fractal space-times}

Geodesic and geodesic deviation equations can be obtained
simultaneously by applying the action principle on the Bazanski
Lagrangian \cite{Ba89, Wa95, Wa99, Ka06}:
\begin{equation}
L= g_{\alpha \beta} U^{\alpha} \frac{D \Psi^{\beta}}{Ds},
\end{equation}
where $D/Ds$ is the covariant derivative. By taking the variation
with respect to the deviation vector ${\Psi^{\rho}}$ and with the
unit tangent vector $U^{\rho}$ one obtains the geodesic equation
and the geodesic deviation equation respectively,
\begin{equation}
\frac{dU^{\alpha}}{ds} +{{\alpha} \brace {\mu \nu}}U^{\mu}U^{\nu}=0 ,\frac{%
D^{2}\Psi^{\alpha}}{Ds^{2}} = R^{\alpha}_{. \beta \gamma \delta}
U^{\beta}U^{\gamma} \Psi^{\delta},
\end{equation}
where ${{\alpha} \brace {\mu \nu}}$ is the Christoffel symbol of
the second kind $R^{\alpha}_{\beta \gamma \delta}$ is the Riemann-
Christoffel curvature tensor. In the case of the fractal
space-times we propose the following form of the Bazanski
Lagrangian $L$,
\begin{equation}
L=g_{\mu \nu }\tilde{V}^{\mu }\frac{\hat{D}_{\pm }\tilde{\Psi}^{\nu }}{Ds}%
+f_{\mu }\tilde{\Psi}^{\nu },
\end{equation}
where $f_{\mu}= (e/m) \tilde{F}_{\mu \nu} V^{\nu} + (1/2m)\tilde{%
R}_{\mu \nu \gamma \delta }S^{\gamma \delta}V^{\nu}$, and the
covariant scale derivatives in the fractal space are defined as
\begin{equation}
\frac{\tilde{D}\tilde{V}}{\tilde{D}s}=\frac{D\tilde{V}^{\alpha }}{Ds}+\tilde{%
V}^{\mu }\cdot D_{\mu }\tilde{V}^{\alpha }-i\mu \left( D^{\mu
}D_{\mu }+\xi R\right) \tilde{V}^{\alpha },
\end{equation}
and
\begin{equation}
\frac{\tilde{D}_{\pm }\left( V^{a}+iU^{a}\right)
}{Ds}=\frac{D\left( V^{a}+iU^{a}\right) }{Ds}+\left( V^{\mu
}+iU^{\mu }\right) \cdot D_{\mu \pm }\left( V^{a}+iU^{a}\right)
-i\mu \left( \partial _{\pm }^{\mu }\partial _{\mu \pm }\right)
\left( V^{a}+iU^{a}\right),
\end{equation}
respectively.

By using the Bazanski approach we can immediately obtain the
Lorentz force equation, the Papapetrou equation (corresponding to
the motion of spinning particles) and the Dixon equation
(describing the motion of spinning particles in electromagnetic
fields as
\begin{equation}
\frac{\hat{D}\tilde{V}^{\mu }}{\hat{D}s}=\frac{e}{m}\tilde{F}_{\nu }^{\mu }%
\tilde{V}^{\nu },\frac{\hat{D}\tilde{V}^{\mu }}{\hat{D}s}=\frac{1}{2m}\tilde{%
R}_{\nu \sigma \rho }^{\mu }\tilde{S}^{\sigma \rho }\tilde{V}^{\nu },\frac{%
\hat{D}\tilde{V}^{\mu }}{\hat{D}s}=\frac{1}{2m}\tilde{R}_{\nu
\sigma \rho }^{\mu }\tilde{S}^{\sigma \rho }\tilde{V}^{\nu
}+\frac{e}{m}\tilde{F}_{\nu }^{\mu }\tilde{V}^{\nu }.
\end{equation}

The quantum covariant derivative in fractal space-times can be
generalized as
\begin{equation}
\frac{\tilde{D}}{\tilde{D}s}=\left[ \tilde{V}^{0}D_{0}-\frac{i\lambda _{c}}{2%
}\left( D^{0}D_{0}+\xi R\right) \right] =0.
\end{equation}
The geodesic equation of motion is
$\tilde{D}\tilde{V}^{0}/\tilde{D}s=0$. By putting $\tilde{V}_{\mu
}=i\lambda D_{\mu }\ln \Psi $, we obtain the Klein-Gordon equation
for a free particle in a curved space as
\begin{equation}
\lambda ^{2}D^{\mu }\ln \Psi D_{\mu }D_{\rho }\ln \Psi
+\frac{\lambda _{c}^{2}}{2}\left( D^{\mu }D_{\mu }D_{\rho }\ln
\Psi +\xi RD_{\rho }\ln \Psi \right) =0.
\end{equation}

The geodesic deviation equation is given by
\begin{eqnarray}
\frac{\hat{D}^{2}\tilde{\Psi}^{\alpha }}{\hat{D}s^{2}}&=&\frac{\hat{D}}{\hat{%
D}s}\left[ \frac{\hat{D}\tilde{\Psi}^{\alpha
}}{\hat{D}s}+\tilde{V}^{\mu }\cdot D_{\mu }\tilde{\Psi}^{\alpha
}-i\mu \left( D^{\mu }D_{\mu }+\xi
R\right) \tilde{\Psi}^{\alpha }\right] =  \nonumber \\
&&\tilde{R}_{\beta \gamma \delta }^{\alpha }\tilde{V}^{\beta }\tilde{V}%
^{\gamma }\tilde{\Psi}^{\delta }=\left( R_{\beta \gamma \delta
}^{\alpha
}+\Xi _{\beta \gamma \delta }^{\alpha }\right) \tilde{V}^{\beta }\tilde{V}%
^{\gamma }\tilde{\Psi}^{\delta }.
\end{eqnarray}

Let us introduce now the covariant derivation operator with
respect to the coordinates as $D\Psi _{\alpha }/Ds=D_{\alpha }\ln
\Phi \left( x,\Psi \right) $ and the covariant derivation operator
with respect to the deviation vector $\tilde{V}^{\alpha }=D_{\Psi
^{\alpha }}\ln \Phi \left( x,\Psi \right) $. Then the geodesic
deviation equation in \ afractal space-time is given by
\begin{eqnarray}
\frac{\hat{D}^{2}D_{\alpha }\ln \Phi }{\hat{D}s^{2}} &=&\frac{\hat{D}}{\hat{D%
}s}\left[ \frac{\hat{D}\left( D_{\alpha }\ln \Phi \right) }{\hat{D}s}%
+D_{\Psi ^{\alpha }}\ln \Phi \cdot D_{\mu }D_{\alpha }\ln \Phi
-i\mu \left(
D^{\mu }D_{\mu }+\xi R\right) D_{\alpha }\ln \Phi \right] =  \nonumber \\
&&\left( R_{\beta \gamma \delta }^{\alpha }+\Xi _{\beta \gamma
\delta }^{\alpha }\right) D^{\Psi ^{\beta }}\ln \Phi D^{\Psi
^{\gamma }}\ln \Phi D^{\Psi ^{\delta }}\ln \Phi .
\end{eqnarray}

\section{Discussions and final remarks}

The theory of scale relativity extends Einstein's principle of
relativity to scale transformations of resolutions, and it gives
up the concept of differentiability of space-time. Its main result
is the reformulation of quantum mechanics from its first
principle. In a fractal space time a small increment of
displacement of the non differentiable four-coordinates along one
of the geodesics can be generally decomposed into its mean and a
fluctuating term. In the present paper we have obtained the path
equations in general scale relativity, as described by a fractal
Riemannian geometry, and we have combined the geodesic equations
and the Schrodinger/ Klein-Gordon equations in a single equation,
which can be reduced to each of them separately if and only if one
uses the averaging procedure and solve the problem of the
integrability of the affine connection due to the curvature of the
space time. Also, we have obtained a quantum analog of the
geodesic deviation equation as defined in a Fractal Space-Time.

\end{document}